\documentstyle[aps,multicol,epsf]{revtex}
\begin{document}
\draft

\title{Why Effective Medium Theory Fails in Granular Materials}

\author{Hern\'an A. Makse$^1$, Nicolas Gland$^{1,2}$,
David L. Johnson$^1$, and Lawrence Schwartz$^1$}
\address{$^1$ Schlumberger-Doll Research, Old Quarry Road, Ridgefield, CT
06877\\
$^2$ Ecole Normale Sup\'erieure, Departement T.A.O., 24 Rue Lhomond, 75005, 
Paris, France}

\date{Phys. Rev. Lett., 13 Dec 1999}
\maketitle
\begin{abstract}

\noindent
Experimentally it is known that the bulk modulus, $K$, and shear
modulus, $\mu$, of a granular assembly of elastic spheres increase
with pressure, $p$, faster than the $p^{1/3}$ law predicted by
effective medium theory (EMT) based on Hertz-Mindlin contact forces.
Further, the ratio $K/\mu$ is found to be roughly pressure independent
but the measured values are considerably larger than the EMT
predictions. To understand the origin of these discrepancies, we have
undertaken numerical simulations of a granular assembly of spherical
elastic grains.  Our results for $K(p)$ and $\mu(p)$ are in good
agreement with the existing experimental data.  We show, also, that
EMT can describe their pressure dependence if one takes into account
the fact that the number of grain-grain contacts increases with $p$.
Most important, the {\it affine} assumption (which underlies EMT), is
found to be valid for $K(p)$ but to breakdown seriously for $\mu (p)$.
This explains why the experimental and numerical values of $\mu (p)$
are much smaller than the EMT predictions.

\end{abstract}

%81.06.Rm: porous materials, granular materials
%81.40.Jj: elasticity, and anelasticity, stress-strain relations

\begin{multicols}{2}

The study of sound propagation and nonlinear elasticity in
unconsolidated granular matter is a topic of great current interest
\cite{review}.  In the simplest experiments, a packing of glass beads
is confined under hydrostatic conditions and the compressional and
shear sound speeds, $v_p$ and $v_s$, are measured as functions of
pressure, $p$ \cite{domenico,yin}.  In the long-wavelength limit, the
sound speeds are related to the elastic constants of the aggregate:
$v_p = \sqrt{(K+4/3\mu)/\rho^*}$ and $v_s = \sqrt{\mu/\rho^*}$, where
$\rho^*$ is the system's density.  In a recent Letter \cite{caroli} 
acoustic measurements were made on bead packs under uniaxial stress 
and it was suggested that long wavelength compresional waves can 
be described in terms of an effective medium.  Thus, it would of great 
value to have a reliable EMT to describe sound propagation as a function
of applied stress.  However, our analysis, together with the work of others, 
raises serious question about the validity of the generally accepted 
theoretical formulation.  The EMT \cite{emt} predicts that
$K$ and $\mu$ both vary as $p^{1/3}$, and that the ratio $K/\mu$ is a
constant (independent of pressure and coordination number) dependent
only on the Poisson's ratio of the material from which the individual
grains are made.

\begin{figure}
\centerline{
\hbox{
\epsfxsize=5.5cm \epsfbox{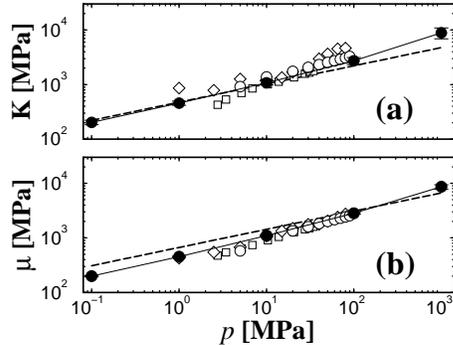}
}
}
\narrowtext
\vspace{.3cm}
\caption{Pressure dependence of the elastic moduli from GD $(\bullet)$, 
experiments [Domenico ($\Box$) \protect\cite{domenico}, Yin ($\Diamond$)
\protect\cite{yin}, and our experiments ($\bigcirc$)], 
and theory (dashed line): (a) bulk modulus, and (b) shear
modulus.}
% We show only the experimental range of pressures for a 
%better comparison. The full GD calculations are shown in Fig. 
%\protect\ref{mf}.}
\label{pressure}
\end{figure}

Experimentally (see Fig. \ref{pressure}),
%\cite{woods},
the bulk and shear moduli increase more rapidly than $p^{1/3}$ and the
values of $K/\mu$ are considerably larger than the EMT prediction.
These discrepancies between theory and experiment could be due to the
breakdown of the Hertz-Mindlin force law at each grain contact as
proposed in \cite{pgg} for the case of metallic beads with an oxide
layer, and in \cite{goddard} for grains with sharp angularities.
Alternatively, they could be associated with the breakdown of some of
the assumptions underlying the EMT, for example,
%that the motion of each
%grain is dictated by the externally applied strain (the affine assumption)
that the number of contacts per grain is pressure independent, which
may not be the case as several authors have suggested
\cite{goddard,jenkins}.

In this Letter we report calculations of $K(p)$ and $\mu(p)$ based on
granular dynamics (GD) simulations using the Discrete Element Method
developed by Cundall and Strack \cite{cundall,wolf}.  Here it is
assumed at the outset that one has an assembly of spherical soft grains
which interact via the Hertz-Mindlin force laws.  We find good
agreement with the existing experimental data, thus confirming the
validity of Hertz-Mindlin contact theory to glass bead aggregates.
Further, we can explain the two problems with EMT described
above. First, if the calculated increase of the average coordination
number with $p$ is taken into account, the modified EMT gives an 
accurate 
description of the bulk modulus found in the simulations, 
$K(p)$; for $\mu(p)$ we
obtain a curve whose shape is in good agreement with the simulation data
but   whose values are
seriously offset therefrom.  Second, the EMT makes the
affine assumption in which the motion of each grain is specified
simply in terms of the externally applied strain (see below).
We show that while the affine assumption is approximately valid for
the bulk modulus, it is seriously in error for the shear modulus; this
is why the EMT prediction of $K/\mu$ differs significantly from the
experimental value.

{\it Numerical Simulations.} At the microscopic level the grains
interact with one another via (1) non-linear Hertz normal forces and
(2) friction generated transverse forces.  The normal force, $f_n$,
has the typical 3/2 power law dependence on the overlap between two
spheres in contact, while the transverse force, $f_t$ depends on both
the shear and normal displacements between the spheres \cite{johnson}.
  For two spherical grains with radii $R_1$ and $R_2$:
\begin{mathletters}
\begin{equation}
f_n = \frac{2}{3}~ C_n R^{1/2}w^{3/2},
\end{equation}
\begin{equation}
\Delta f_t = C_t (R w)^{1/2} \Delta s.
\end{equation}
\end{mathletters}
\noindent
Here $R=2 R_1 R_2/(R_1+R_2)$, the normal overlap is $w=
(1/2)[(R_1+R_2) - |\vec{x}_1 - \vec{x}_2|]>0$, and $\vec{x}_1$,
$\vec{x}_2$ are the positions of the grain centers.  The normal force
acts only in compression, $f_n = 0$ when $w<0$.  The variable $s$ is
defined such that the relative shear displacement between the two
grain centers is $2s$. The prefactors $C_n=4 G / (1-\nu)$ and $C_t = 8
G / (2-\nu)$ are defined in terms of the shear modulus $G$ and the
Poisson's ratio $\nu$ of the material from which the grains are made.
In our simulations we set $G=29$ GPa and $\nu = 0.2$.
%; the typical
%values for glass, there being little variation among different
%silicate glasses at room temperature.
We assume a distribution of
grain radii in which $R_1=0.105$ mm for half the grains and
$R_2=0.095$ mm for the other half.  Our results are, in fact, quite
insensitive to the choice of distribution.  We also include a viscous
damping term to allow the system to relax toward static equilibrium
\cite{friction}.

\vspace{-.1cm}
\begin{figure}
\centerline{
\hbox{
\epsfxsize=5.5cm \epsfbox{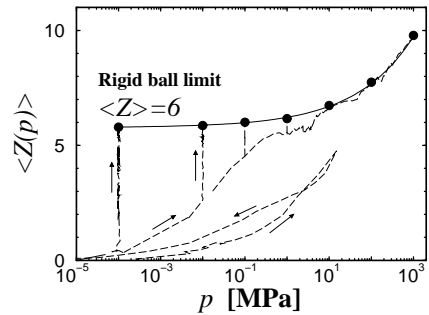}
}
}
\caption{
Coordination number versus pressure. The dashed line is a fit 
according
to Eq. \protect\ref{Z}.}
\label{coord}
\end{figure}

Our calculations begin with a numerical protocol designed to mimic the
experimental procedure used to prepare dense packed granular
materials.  In the experiments the initial bead pack is subjected to
mechanical tapping and ultrasonic vibration in order to increase the
solid phase volume fraction, $\phi_s$.  The simulations begin with a
{\it gas} of 10000 spherical particles located at random positions in
a periodically repeated cubic unit cell approximately 4 mm on a side.
At the outset, the transverse force between the grains is turned off
($C_t = 0$).  The system is them compressed slowly until a specified
value of $\phi_s$ is attained (see dashed lines in Fig. \ref{coord}).
The compression is then stopped and the grains are allowed to relax.
If the compression is stopped before reaching the critical volume
fraction, $\phi_s\sim 0.64$, corresponding to random close packing
(RCP), the system will relax to zero pressure and zero coordination
number, since the system cannot equilibrate below RCP.  The
compression is then continued to a point above the critical packing
fraction and the target pressure is maintained with a ``servo''
mechanism \cite{cundall} which constantly adjusts the applied strain
until the system reaches equilibrium.  Because there are no transverse
forces, the grains slide without resistance during the relaxation
process and the system reaches the high volume fractions found
experimentally.

The simulated granular aggregate
relaxes to equilibrium states in which the average coordination
number, $<Z (p)>$, increases with pressure as seen in
Fig. \ref{coord}.  For low pressures compared with the shear modulus
of the beads $<Z>\approx 6$, while in two dimensions the same preparation
protocol gives $<Z>\approx 4$.  Such low coordination numbers can be
understood in terms of a constraint argument for frictionless rigid
balls \cite{edwards,alexander}, which gives $<Z>=2 D$, where D is the
dimension.  These values should be valid in the limit of low pressure
when the beads are minimally connected near RCP
\cite{alexander} (or in the rigid ball limit $G\to\infty$).
For large values of the confining pressure more grains are brought
into contact, and the coordination number increases \cite{robust}.  
Empirically, we
find that
\begin{equation}
<Z(p)> = 6 + \left(\frac{p}{0.06 ~\mbox{MPa}}\right)^{1/3}.
% 5.775 + 0.3161342 log p  -0..3505121
\label{Z}
\end{equation}

{\it Comparison with Experiment}. Consider, now, the 
calculation of the elastic moduli of the system as
function of pressure.  Beginning with the equilibrium state describe
above, we first restore the transverse component of the contact force
interaction.  We then apply a small perturbation to the system and
measure the resulting response. The shear modulus is calculated in two
ways, from a pure shear test, $\mu = (1/2) \Delta \sigma_{12}/\Delta
\epsilon_{12}$, and also from a biaxial test, $\mu = (\Delta
\sigma_{22}-\Delta \sigma_{11} )/2 (\Delta \epsilon_{22}-\Delta
\epsilon_{11})$.  The bulk modulus is obtain from a uniaxial
compression test, $K+4/3\mu= \Delta \sigma_{11}/\Delta \epsilon_{11}$.  
Here the stress, $\sigma_{ij}$, is determined from the
measured forces on the grains \cite{cundall}, 
and the strain, $\epsilon_{ij}$, is
determined from the imposed dimensions of the unit cell.

\vspace{-.5cm}

In Fig. \ref{pressure} our calculated values of the elastic moduli as
a function of pressure are compared with EMT and with experimental
data.  Because there is a considerable degree of scatter in the
experimental results we performed our own experiments with standard sound
propagation techniques. A set of high quality glass beads of diameter
$45 \mu$m are deposited in a flexible container of 3 cm height and 2.5
cm radius. Transducers and a pair of linear variable differential
transformers (for measurement of displacement) were placed at the top
and bottom of the flexible membrane, and the entire system was put
into a pressure vessel filled with oil.  Before starting the
measurements, a series of tapping and ultrasonic vibrations were
applied to the container in order to let the grains settle into the
best possible packing.  We then applied confining pressures ranging
from 5 MPa to 100 MPa.  The pressure was cycled up and down several
times until the system exhibited minimal hysteresis.  At this point
shear and compressional waves were propagated by applying pulses with
center frequencies of 500 KHz.  The sound speeds and corresponding
moduli were obtained by measuring the arrival time for the two sound
waves.

From Fig. \ref{pressure} we see that our experimental and numerical
results are in reasonably good agreement.  Also shown are measured
data from Domenico \cite{domenico} and Yin \cite{yin}.  Clearly, the
experimental data are somewhat scattered.  This scatter reflects the
difficulty of the measurements, especially at the lowest pressures
where there is significant signal loss.  Nevertheless, our calculated
results pass through the collection of available data.
 Also shown in Fig. \ref{pressure} are the EMT predictions \cite{emt}
\begin{mathletters}
\begin{equation}
K= \frac{C_n}{12 \pi} 
~ (\phi_s Z)^{2/3} ~\left(\frac{6 \pi p}{C_n}\right)^{1/3},
\end{equation}
\begin{equation}
\mu= \frac{C_n + (3/2) C_t}{20 \pi} ~  (\phi_s Z)^{2/3}
~\left(\frac{6 \pi p}{C_n}\right)^{1/3}.
\end{equation}
\label{EMT}
\end{mathletters}
The EMT curves are obtained using the same parameters as in the
simulations; we also set $Z=6$ and $\phi_s=0.64$, independent of pressure.
At low pressures we see that $K$ is well described by EMT.  At larger
pressures, however, the experimental and numerical values of $K$ grow
faster than the $p^{1/3}$ law predicted by EMT.  The situation with
the shear modulus is even less satisfactory.  EMT overestimates $\mu
(p)$ at low pressures but, again, underestimates the increase in $\mu
(p)$ with pressure.

\begin{figure}
\centerline{
\hbox{
\epsfxsize=5.5cm \epsfbox{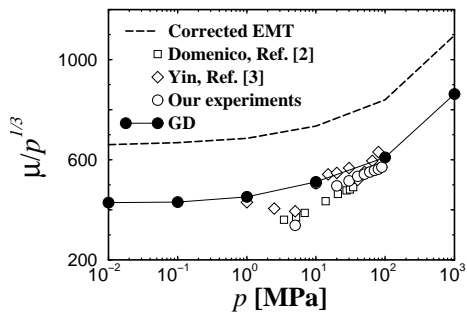}
}
}
\caption{Shear modulus and corrected EMT taking into account the
pressure  dependence of $<Z(p)>$ from Fig. \protect\ref{coord} as well as
$\phi_s(p)$.}
\label{mf}
\end{figure}
\vspace{-.3cm}
To investigate the failure of EMT in predicting the correct pressure
dependence of the moduli, we plot in Fig. \ref{mf} the shear modulus
divided by $p^{1/3}$. For such a plot, EMT predicts a horizontal
straight line but we see that the numerical and experimental results
are clearly increasing with $p$.  We can explain this behavior by
modifying Eq. (3b) to take account of the pressure dependence of the
coordination number $<Z(p)>$ from Fig. \ref{coord} and the pressure
dependence of $\phi_s(p)$ (which is a much smaller effect).  The
modified EMT is also plotted in Fig. \ref{mf} and we see that it
predicts the same trend with pressure as the simulations.  The
experimental data also seem to be following this trend but more data
over a larger pressure range are clearly needed.  Not shown in
Fig. \ref{mf} is a similar analysis of $K(p)$ but the result is that
the modified EMT is in essentially exact agreement with our numerical
simulations.  It is for this reason that we focus on $\mu(p)$.

Another way of seeing the breakdown of EMT is to focus on the ratio $K
/ \mu$.  According to Eqs. (\ref{EMT}), $K / \mu =5/3
(2-\nu)/(5-4\nu)$, independent of pressure, a value which depends only
on the Poisson's ratio of the bead material.  The experiments give
$K/\mu\approx 1.1-1.3$. Our simulations give $K/\mu\approx 1.05\pm0.1$
in good agreement with experiments.  EMT predicts $K/\mu=0.71$, if we
take $\nu=0.2$ for the Poisson's ratio of glass. [The EMT
 prediction is
quite insensitive to variations of $\nu$; $K/\mu = 0.71 \pm 0.04$ for
$\nu = 0.2 \pm 0.1$.]

\begin{figure}
\centerline{
\hbox{
\epsfxsize=5.cm \epsfbox{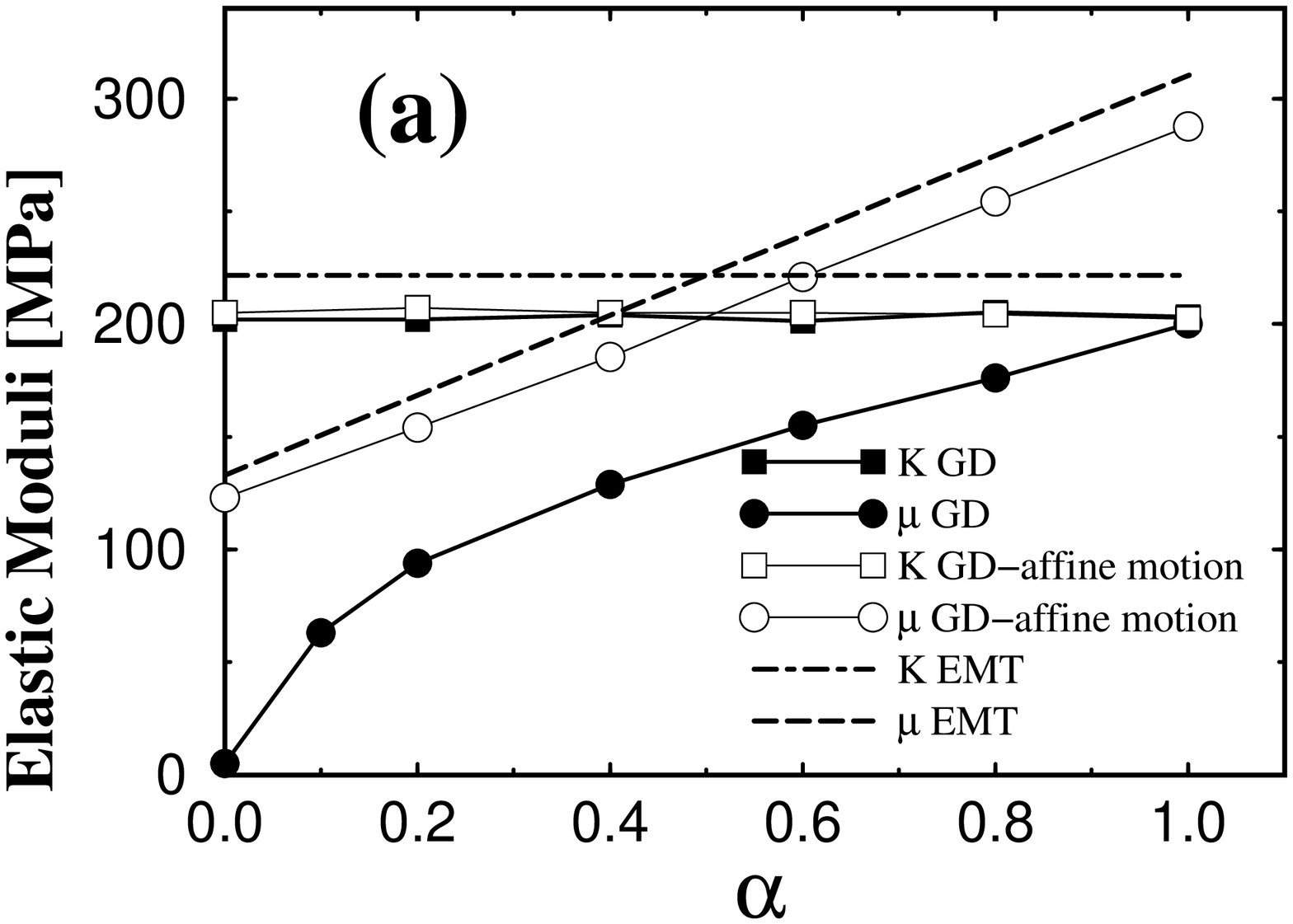}
}
}
\centerline{
\hbox{
\epsfxsize=5.cm \epsfbox{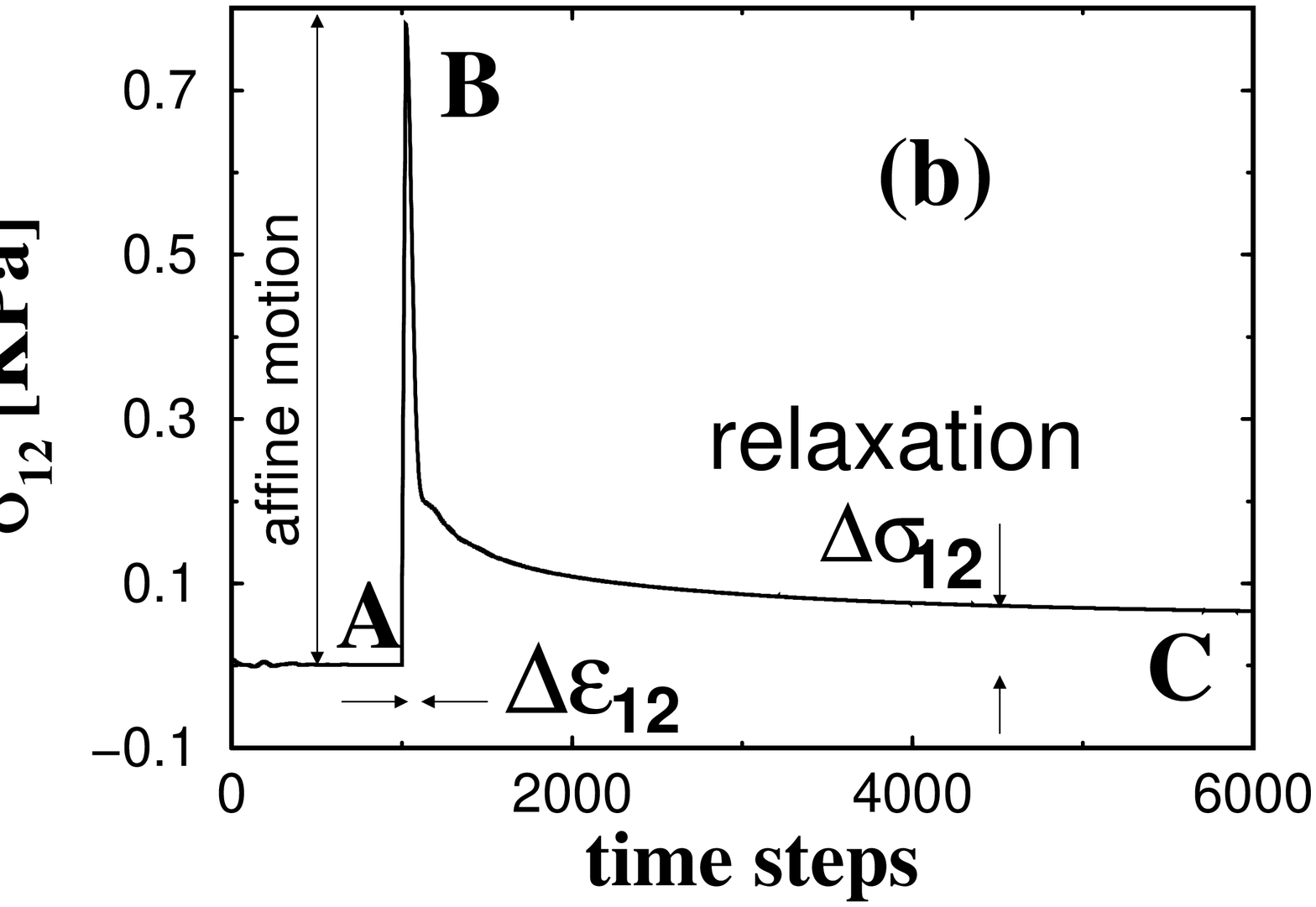}
     }
}
\caption{(a) $K(\alpha)$, and $\mu(\alpha)$ versus $\alpha$ for a fixed
$p=100$  KPa. (b) Relaxation of the shear stress (B$\to$C)
after an affine motion (A$\to$B) in the calculation of the shear modulus.}
\label{alpha}
\end{figure}

To understand why $\mu$ is over predicted by EMT we must examine the
role of transverse forces and rotations in the relaxation of the grains.
[These effects do not play any role in the calculation of the bulk
modulus.]  Suppose we re-define the transverse force by introducing a
multiplicative coefficient $\alpha$, viz: $\Delta f_t = \alpha C_t
(Rw)^{1/2} \Delta s$; with $\alpha=1$ we recover our previous results.
To quantify the role of the transverse force on the elastic moduli, we
calculate $K(\alpha)$ and $\mu(\alpha)$ at a given pressure $p=100$
KPa [Fig. \ref{alpha}a].  This pressure is low enough that the changing
number of contacts is not an issue.  Surprisingly, the shear modulus
becomes negligibly small as $\alpha \rightarrow 0$.  As expected, $K$
is independent of the strength of the transverse force.  To compare
with the theory we also plot the prediction of the EMT,
Eq. (\ref{EMT}) in which $C_t$ is rescaled by $\alpha C_t$.  We see
that the EMT fails in taking into account the vanishing of the shear
modulus as $\alpha\to0$.  However it accurately predicts the value of
the bulk modulus, which is independent of $\alpha$.

There are two main approximations in the EMT: (1) All the spheres 
are statistically the same, and it is assumed that there is 
an isotropic distribution of contacts around a given sphere.  (2) An 
affine approximation is used, i.e., the spheres at position $X_{j}$
are moved a distance $\delta u_i$ in a time interval $\delta t$
according to the macroscopic strain rate $\dot{\epsilon}_{ij}$ by
$\delta u_i=\dot{\epsilon}_{ij} X_{j} \delta t$.  The grains are
always at equilibrium due to the assumption of isotropic distribution
of contacts and further relaxation is not required.

In the GD calculation of the shear modulus an affine perturbation is
first applied to the system.  The shear stress increases (from A to B
in Fig. \ref{alpha}b) and the grains are far from
equilibrium since the system is disordered.  
The grains then relax towards equilibrium (from B to C),
and we measure the resulting change in stress from which the modulus
is calculated.  To better understand the approximations involved in
the EMT, suppose we repeat the GD calculations taking into account
only the affine motion of the grains and ignoring the subsequent
relaxation.  The resulting values of the moduli are plotted in
Fig. \ref{alpha}a as open symbols and we see that the moduli calculated
this way are very close to the EMT predictions.  Thus, the difference
between the GD and EMT results for the shear modulus lies in the
non-affine relaxation of the grains; this difference being largest
when there is no transverse force.  By contrast, grain relaxation
after an applied {\it isotropic} affine perturbation is not
particularly significant and the EMT predictions for the bulk modulus
are quite accurate.

\vspace{-.2cm}
The surprisingly small values of $\mu$ found as $\alpha \rightarrow 0$
can be understood as a melting of the system that occurs when the
system is close to the RCP fraction. This fluid like behavior (when
$C_t = 0$) is closely related to the melting transition seen in
compressed emulsions \cite{grest} and foams \cite{durian}.  At the RCP
fraction the system behaves like a fluid with no resistance to shear.
By contrast, molecular dynamics simulations of glasses, in which the
atoms interact by purely longitudinal forces, predict non-vanishing
shear speeds \cite{GNR}.  The crucial difference between these two
systems is the local coordination of the particles.  In the granular
system, the coordination number near RCP (where the balls are only
weakly deformed) is $<Z> = 6$; the system is at its minimal
coordination number. In glasses, however, the number of neighbors is
closer to 10 and the motion of the grain is highly constrained.

{\it Conclusions}. Our GD simulations are in good agreement with the
available experimental data on the pressure dependence of the elastic
moduli of granular packings.  They also serve to clarify the
deficiencies of EMT.  Grain relaxation after an infinitesimal affine
strain transformation is an essential component of the shear (but not
the bulk) modulus.  This relaxation is not taken into account 
in the EMT.  In the limit $\alpha \to 0$ a packing of nearly rigid particles
responds to an external isotropic load with an elastic 
deformation and a finite $K$. By contrast, such a system
cannot support a shear load ($\mu \to 0$) without severe
particle rearrangements.  This may indicate a ``fragile'' 
state of the system \cite{fragile} where  inter particle
forces are organized along ``force chains''
(stress paths carrying most of the forces in the system) oriented along 
the principal stress axes. Such fragile networks support, elastically, 
only perturbations compatible with the structure of  force
chains and deform plastically otherwise.  Clearly, there is a 
need for an improved EMT; recent work on stress
transmission in minimally connected networks
\cite{edwards,alexander} may provide an alternative formulation
and allow to describe properly the  response of granular
materials to perturbations.

ACKNOWLEDGMENTS.
We would like to thank J. Dvorkin,
J. St. Germain, B. Halperin, J. Jenkins, 
and D. Pissarenko for many stimulating discussions.

%\narrowtext

%\vspace{-.9cm}

\end{multicols}


\begin{references}
\bibitem{friction} When the $f_t$ exceeds the Coulomb threshold, 
$\mu_c f_n$, the grains slide and $f_t = \mu_c f_n$, where $\mu_c$ is the
coefficient of kinetic friction between the spheres.  In fact, since we work
with infinitesimal distortions, slippage never occurs.
\bibitem{robust}
To test the robustness of these results, we have employed a 
second protocol in which the system is prepared by compressing
to a point beyond the RCP fraction then letting the grains relax 
to equilibrium without the servo mechanism.  The final $<Z(p)>$ 
curve is essentially identical to the one shown in Fig. 
\protect\ref{coord}.
\end{references}
\end{document}